%% file: covi_paper.tex
\documentclass{llncs}
\usepackage{pslatex}
\usepackage[english]{babel}
\usepackage{graphicx,multirow}
\usepackage{amsmath}
\usepackage{amssymb}
\usepackage{url}
\usepackage{comment}
\usepackage{algorithm}
\usepackage{algorithmic}
\usepackage{booktabs}
\usepackage{xspace}
\usepackage{xcolor}
\usepackage{colortbl} 

\usepackage{subfig}

\newcommand{\fyle}[1]{{\texttt{#1}}}
\newcommand{\namelabel}[1]{{\textsc{#1}}}
\newcommand{\vmit}[1]{\mbox{\textit{#1}}}

\newcommand{\LOUDS}{{\namelabel{LOUDS}}}
\newcommand{\COvI}{{\formatfonc{COvI}}}
\newcommand{\covi}{{\formatfonc{COvI}}}
\newcommand{\AC}{{\formatfonc{Full AC}}}
\newcommand{\fullac}{{\formatfonc{Full AC}}}
\newcommand{\BP}{{\namelabel{BP}}}

\newcommand{\method}[1]{{\textsf{#1}}}

\newcommand{\D}{\hphantom{0}}

\newcommand{\ie}{\emph{i.e.}}
\newcommand{\lgr}[1]{\vert #1 \vert}
\newcommand{\gb}[1]{{\mbox{$#1$~GB}}}

\newcommand{\formatfonc}[1]{\ensuremath{\mathtt{#1}}\xspace}
\newcommand{\rank}{\formatfonc{rank}}
\newcommand{\select}{\formatfonc{select}}

\newcommand{\eg}{\emph{e.g.}}

\newcommand{\zero}{\mathtt{0}}
\newcommand{\un}{\mathtt{1}}
\newcommand{\zu}{\left\{ \zero, \un \right\}}

\newcommand{\ov}{\mathtt{ov}}

\newcommand{\pair}[1]{\langle #1 \rangle}

\usepackage{hyperref}

\hypersetup{ 
  pdfauthor   = {Rodrigo Canovas, Bastien Cazaux and Eric Rivals},%
  pdftitle    = {The Compressed Overlap Index},%
  pdfsubject  = {algorithms},%
  pdfkeywords = {data structure, query, stringology, Aho-Corasick, overlap graph, compression},%
  pdfcreator  = {PDFLaTeX},%
  pdfproducer = {PDFLaTeX},
  pagebackref = false,
  colorlinks = true,
  citecolor = blue,
  hyperfigures = true,
  pdftex
}

\title{The Compressed Overlap Index\thanks{This work is supported by the \href{http://www.ibc-montpellier.fr}{Institut de Biologie Computationnelle} (ANR-11-BINF-0002) and D\'efi {MASTODONS C3G} from CNRS.}}

\author{Rodrigo C\'anovas \inst{1,2} \and Bastien Cazaux\inst{1,2} \and Eric Rivals\inst{1,2}}
\institute{
  L.I.R.M.M. 
  Universit\'e de Montpellier, CNRS, U.M.R. 5506, Montpellier, France\\
  \and
  Institut Biologie Computationnelle, Montpellier, France
  \\\email{canovas-ba@lirmm.fr}, \email{cazaux@lirmm.fr}, \email{rivals@lirmm.fr}
  }

\begin{document}
\maketitle

\begin{abstract}
For analysing text algorithms, for computing superstrings, or for testing random number generators, one needs to compute all overlaps between any pairs of words in a given set. The positions of overlaps of a word onto itself, or of two words, are needed to compute the absence probability of a word in a random text, or the numbers of common words shared by two random texts. In all these contexts, one need to compute or to query overlaps between pairs of words in a given set.
For this sake, we designed {\covi}, a compressed overlap index that supports multiple queries on overlaps: like computing the correlation of two words, or listing pairs of words whose longest overlap is maximal among all possible pairs. {\covi} stores overlaps in a hierarchical and non redundant manner. We propose an implementation that can handle datasets of millions of words and still answer queries efficiently. Comparison with a baseline solution -- called {\fullac} -- relying on the Aho-Corasick automaton shows that {\covi} provides significant advantages. For similar construction times, {\covi} requires half the memory {\fullac}, and still solves complex queries much faster. 

\end{abstract}

\section{Introduction}\label{sec:intro}
\input{new_intro}

\subsection{Basic Concepts}\label{sec:concepts}
\input{basics}

\section{Compressed Overlap Index}\label{sec:coi}
\input{coi}

\section{Experiments}\label{sec:exp}
\input{experi}

\section{Conclusions}\label{sec:conclusion}

Here, we introduced {\COvI}: to our knowledge, the first available compressed representation of an overlap index. 
It exploits a reduced version of Aho-Corasick automaton, and can store all (maximal and non maximal) overlaps 
between any pair of input words in linear space. {\COvI} offers multiple types of queries: on overlaps between 
two words, or between one word onto any other word, and a query for retrieving the pairs of words having the 
longest overlap over all possible pairs.
Experiments  showed the performance of {\COvI} both in terms of construction time, of space usage, and of querying 
times for different kinds of texts and alphabets. It outperforms a solution based on a version of the Aho-Corasick 
automaton.
%
%
Future work include improving some query algorithms (\eg, $\formatfonc{max\text{-}ov}(x,y)$) to reach a linear complexity, creation of queries addressing a subset of the input words, or making {\COvI} dynamic to allow insertions/deletions of words without requiring reconstruction from scratch. We also consider studying how to implement and modify a recent work of G. Manzini~\cite{Man16}, which may lower memory usage but increase query times. In the same line, it would be interesting to assess {\COvI} performance in large scale experiments, as in \cite{SD10}.

An interesting perspective is to use {\COvI} to implement (approximation) algorithms for the shortest superstring problem 
(and its variants), and to test their scalability and their ability to solve this difficult question for various inputs and 
alphabet sizes.

Finally, all implementations from this article are publicly available at \url{https://github.com/rcanovas}.
\bibliography{covi_ref}
\end{document}

%% file: new_intro.tex

A text, a word, or a string $u$ is a sequence of letters taken from an alphabet $\Sigma$. Given two words $u, v$ over $\Sigma$, $u$ overlaps $v$ if a suffix of $u$ equals a prefix of $v$. Overlaps between the words of a given set are crucial in numerous applications: for computing word statistics or superstrings, for text compression, for analysing text algorithms, or for testing random number generators. In those various contexts, it is valuable to offer a versatile and scalable solution to compute such sets of overlaps. Here, we propose a data structure, dubbed {\covi}, which computes and indexes the overlaps of a set of words, and can thus be later queried to obtain the desired overlaps. Before, exposing our solution, let us dwell on the motivations for computing overlaps, starting with application in word statistics.

Consider a finite Bernoulli random text $T$ over $\Sigma$ (i.e., with independent and identically distributed symbols), and two words $u$ and $v$ of the same length over $\Sigma$. Do $u$ and $v$ have the same probability of occurring in $T$? In general, the answer is no since it depends on the self-overlaps of each word. For instance over $\Sigma = \{0,1\}$, the words $u := 00$ and $v := 01$ have respectively a probability of $3/8$ and of $1/2$ of occurring in a text of length $3$. Moreover, $u$ can have two occurrences in such a text, while $v$ cannot. The difference is due to the fact that occurrences of $u$ can overlap themselves, while those of $v$ cannot. More generally, the absence probability, the waiting time of a word (the number of symbols before the first occurrence), the return probability (the number of positions between successive occurrences), the total number of occurrences in a random text of length $n$, \textbf{all} depend on the self-overlaps of this word \cite{RD99}. All self-overlap positions can be encoded in a single binary vector called the autocorrelation: for instance the autocorrelation of $u:= \text{abracadabra}$ is $c(u):= 10000001001$, where a $1$ denotes a self-overlap (see \cite{GuOd81} for a definition). In \cite{SED:FLA:1996}, derives a formula for the absence probability in function of the autocorrelation. Such statistics are useful many contexts, for instance to evaluate the significance of finding a word in a collection of text (in Information retrieval) or of finding that many occurrences of a DNA binding motif in a genome sequence (in Bioinformatics).  

Now, because of possible overlaps, the occurrences of distinct words in a text are also interdependent.  Like for a single word, when studying a set of words, their occurrences can overlap, and this influences their probability of occurring together or being absent together in random texts. These probabilities depend on the mutual overlaps of the words \cite{Rahmann-master-thesis}, which are encoded in binary vectors called correlations.\\
\textbf{Example}
For the two words u := atatat, v := tggata over $\Sigma :=\{a,c,g,t\}$, their autocorrelations are
c(u) := 101010, c(v) = 100000, and their correlations are c(u,v) = 000001, and c(v,u) = 000101, where c(x,y) denotes the correlation of x over y, that is the binary encoding of the position where $x$ overlaps $y$ from the left. Note that with this notation, c(u) = c(u,u). 

 Such probabilities, whose computation requires to list the correlations of pairs of words in the set, are used in many contexts. For instance, the number of common words of length $q$ (shared by two texts) serves to approximate the distance or the similarity between texts \cite{Jokinen-Ukkonen-MFCS91,QUASAR-RECOMB99}, and co-occurrence probabilities are then employed to optimise filtration criteria in similarity search algorithms\cite{RR2000,RahmannRivalsCPC}. Last, such word statistics are heavily used for testing random number generators~\cite{Leopardi:2009}, when considering the set of all possible words of length $q$ (which is a very specific case). 

In \cite{RahmannRivalsCPC}, the computation of correlation vectors for a set of words is listed as an interesting open problem, which was solved by hand for small cases. An algorithm for the special case of a set containing all possible words of length $q$ was proposed in ~\cite{Leopardi:2009}. Hence, the goal of {\covi} is to provide a general, versatile, and scalable  tool to compute word correlations for any set of words.

In bioinformatics, computation of superstrings models the question of DNA assembly, which aims at inferring a target DNA sequence from a set of short overlapping DNA fragments (also called reads). However, in practice, one considers approximate overlaps because of sequencing errors, and also overlaps between reverse complementary fragments since the DNA is double stranded. Specialised, efficient data structures have been designed to compute the longest overlap for any read pair and to represent them in a graph (e.g. \cite{Simpson-Durbin-sga-2012}). A theoretical model is the \emph{Overlap Graph}: a complete, weighted digraph in which each input sequence is a node, and the arc linking two nodes is weighted by the length of their longest overlap. {\covi} computes a data structure that subsumes the Overlap Graph, but requires much less space, and can be queried to build the Overlap Graph.

Below, we introduce known data structures required for the construction of {\covi}, whose construction and overlap queries are explained in Section~\ref{sec:coi}.  We investigate the performance of {\covi} in Section~\ref{sec:exp}, and conclude in Section~\ref{sec:conclusion}

%


%% file: basics.tex
Let $\Sigma$ denote a finite alphabet of cardinality $\sigma$. A string or word $u$ over $\Sigma$ is a 
sequence of characters from $\Sigma$ of length $\lgr{u}$. For any integer $1 \leq i \leq j \leq \lgr{u}$, we denote the the $i$-th character of $u$ by $u[i]$, and the substring comprised between positions $i$ and $j$ by $s[i,j]$. A substring of $u$ is a prefix (resp. suffix) if it starts at position $1$ (resp. ends at position $\lgr{u}$). A suffix  (resp. prefix) of $u$ is said \emph{proper} if it differs from $u$.
Now we define the non symmetrical notion of overlaps between two strings. Let $u,v$ be strings over $\Sigma$. $u$ overlaps $v$ if a non empty suffix of $u$ equals a prefix of $v$ (\ie, there exists an integer $k$ such that $u[\lgr{u}-k+1, \lgr{u}] = v[1,k]$). Then the string $v[1,k]$ is a \emph{right overlap} for $u$ or a \emph{left overlap} for $v$. The longest overlap of the pair $u$ onto $v$ is denoted $\ov(u,v)$. 

Throughout this work, the input consists in a set $P := \{s_1, \ldots, s_p\}$ of $p$ finite words. 
Then, we denote by $Ov^+(P)$ the set of all overlaps between any two words of $P$.

\subsubsection{Trie}
\begin{figure}[t]
  \begin{center}
    \includegraphics[scale=0.14]{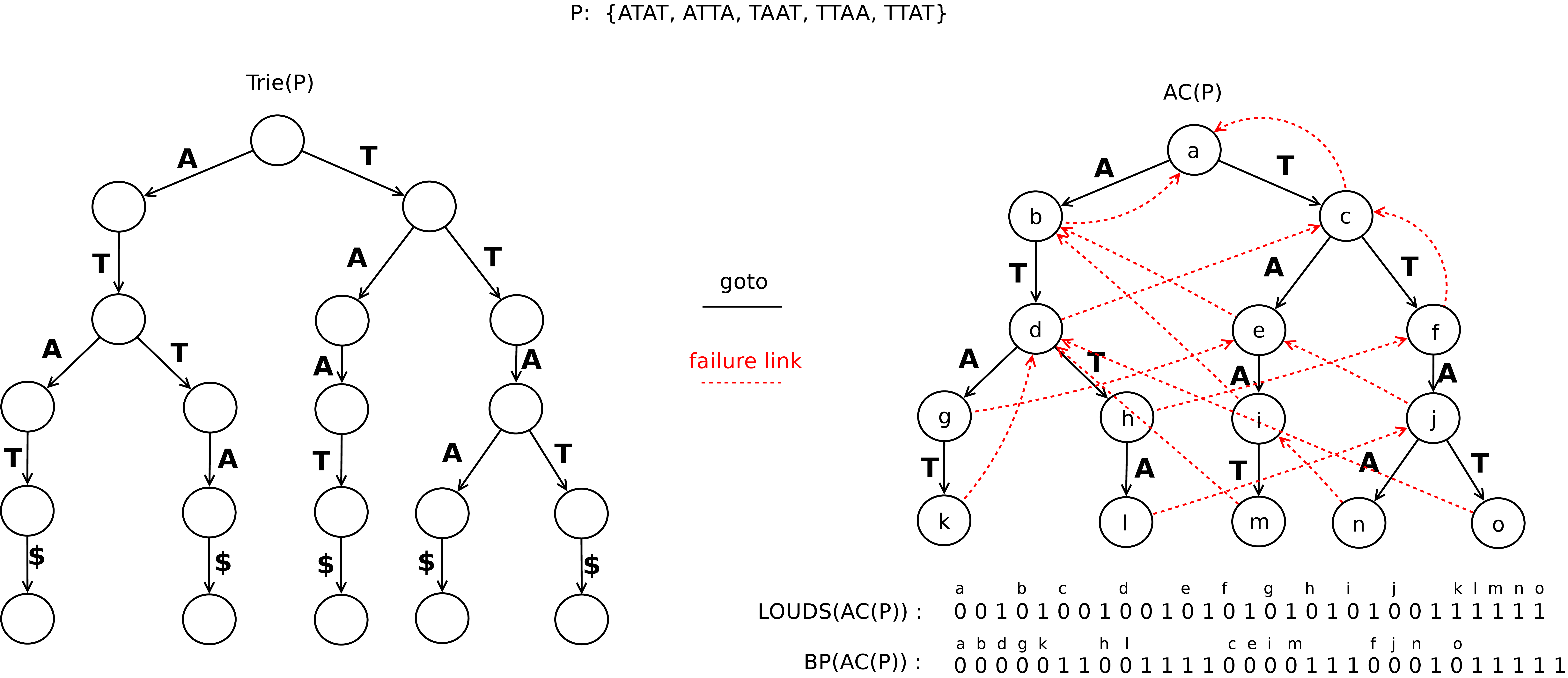} 
  \end{center}
  \caption{Running example with $P = \{ATAT, ATTA, TAAT, TTAA, TTAT\}$. Left: the \vmit{trie} of $P$. Right: Aho-Corasick automaton ($AC$) of $P$, with the {\LOUDS} and {\BP} representations of the $AC$  tree topology.
}
  \label{fig:trie-ac}
\end{figure}

Consider $P$ and assume that no word of $P$ is prefix of another (which can be achieved by appending a special symbol to each word). The \vmit{trie} is a tree, labelled on its arcs, designed to store a set of words \cite{Knu73} (see Figure~\ref{fig:trie-ac}, which is a running example). It has $p$ leaves and it spells out each word $s_i$ on a distinct branch ($s_i$ equals the  concatenation of the labels from the root to its leaf). Each node $v$ uniquely represents a distinct prefix of the words in $P$, with the root being the empty prefix. Hence, two nodes share a common ancestor node $v$ if the string $v$ is a common prefix of both.

\subsubsection{Aho-Corasick Automaton}
In $1975$, Aho \& Corasick proposed the first algorithm solving the exact set pattern matching problem, which for a set $P$ and text $T$ finds all the occurrences of words of $P$ within $T$~\cite{AC75,G97}. Their algorithm builds an automaton for $P$ and then processes $T$ using it -- see Figure~\ref{fig:trie-ac} for the  Aho-Corasick automaton ($AC$) of our running example.  In $AC$, the \vmit{states} are exactly the nodes of the trie $T$ of $P$, and the \vmit{goto} transitions are the arcs of $T$, which link a node to its children.  Hence, to a state $s$ is associated to the prefix represented by the corresponding node of $T$. The AC also includes \vmit{failure} links. The \vmit{failure} link of state $s$ (or simply of $s$) points to state $r$ if and only if $r$ represents the longest proper suffix of $s$ among all the states. Hence, one gets a characterisation of nodes representing overlaps:
\begin{proposition}[\cite{Ukkonen-linear-greedy-1990}]\label{prop-overlap-node}
  Let $s,r$ be two nodes/states in $AC(P)$.
  If $r$ is the node pointed by the failure link of node $s$, and $w$ is a leaf in the subtree of $r$ in $T$, then $r$ is an overlap from the string represented by $s$ onto the string represented by $w$ (\ie, $s$ overlaps $w$).
\end{proposition}
If $s$ is a leaf, we can reach the nodes representing all its overlaps with over words of $P$ by following the chain of failure links starting in $s$ and ending at the root of $T$. Conversely, nodes that are not on such chains (starting at leaves) are not overlaps between words of $P$.

\subsubsection{Rank and Select}
To build {\COvI}, we use binary vectors that support \rank and {\select} operations.  Given a sequence $S$ over $\zu$, and a position $i$ such that $1 \leq i \leq \lgr{S}$, then for any symbol $a$ in $\zu$, $\mbox{\rank}_a(S,i)$ returns the number of occurrences of $a$ until position $i$ in $S$, while ${\select}_a(S,i)$ finds the position in $S$ of the $i$-$th$ occurrence of $a$.

For a binary vector (also called bitmap) of length $n$, it is possible to answer {\rank} and {\select} in constant time using $n+o(n)$ bits \cite{Mun96,Cla98} ($n$ bits for storing the bitmap itself and $o(n)$ additional bits for the data structures).  In practice, such structure requires $5\%$ extra space over the bitmaps size~\cite{GGMN05}.

\subsubsection{Tree Representation} 

The topology of a general tree $\cal T$ of $n$ nodes can be represented succinctly 
using $2n$ bits, and by adding the support of {\rank} and {\select}, one can support 
many tree operations in constant time ~\cite{ACNS14}. In this work we used two different 
succinct tree representations: {\LOUDS} and {\BP}. 

The level-ordered unary degree sequence or {\LOUDS} ~\cite{BDMRRR05,Jac89} is built 
by traversing the nodes of $\cal T$ in level-wise order, and for a node with $i$ children, 
we write $i$ zeros and a one into the bitmap. 
The balanced parentheses sequence or {\BP} ~\cite{Jac89,MR01} is built from a depth-first
preorder traversal of $\cal T$, writing an zero (opening parenthesis) when arriving to a 
node for the first time, and a one (closing parenthesis) when going up (after traversing 
the subtree of the node).  Figure~\ref{fig:trie-ac} shows an example of {\LOUDS} and {\BP}. 

Experiments have shown that {\LOUDS} is very efficient when only the simplest operations 
are needed, like accessing a node's parent or going to an arbitrary child of a node, 
while {\BP} also supports in constant time complex operations, such as computing the depth of a node~\cite{ACNS14}. 
%


%% file: coi.tex

Here, we explain the algorithm to build the index {\covi}, and the procedures that implement the following queries. Let $x,y$ be two strings of $P$, and $q$ be a positive integer.
\begin{itemize}
\item $\formatfonc{correlation}(x,y)$: gives $c(x,y)$, \ie\ the correlation of $x$ over $y$.  
\item $\formatfonc{max\text{-}ov}(x,y)$:  gives the length of the maximum overlap from $x$ to $y$ (\ie,  $\lgr{\ov(x,y)}$).

\item $\formatfonc{all\text{-}right\text{-}ov}(x)$: gives an array of size $\lgr{P}$ containing $\lgr{\ov(x,z)}$ for all $z$ in $P$.
\item $\formatfonc{all\text{-}left\text{-}ov} (y)$: gives an array of size $\lgr{P}$ containing $\lgr{\ov(z,y)}$ for all $z$ in $P$.

\item $\formatfonc{global\text{-}max\text{-}ov()}$: gives all words $x$ in $P$, such that there exists $z_x \in P$ such that  $\lgr{\ov(x,z_x)} = max\{\lgr{\ov(w,z)} \text{ for all } w, z \in P\}$. It gives the words having the longest possible maximum overlaps among all possible pairs of $P\times P$.

\item $\formatfonc{threshold\text{-}right\text{-}ov}(x,q)$: gives an array of pairs $\pair{z,l}$ where $z \in P$,  $\lgr{\ov(x,z)} = l$ and $l > q$. The query $\formatfonc{threshold\text{-}left\text{-}ov}(x,q)$ is defined similarly.
\end{itemize}

The queries $\formatfonc{correlation}$ and $\formatfonc{max\text{-}ov}$ works for a pair of words, while queries $\formatfonc{all\text{-}right\text{-}ov}(x)$ works of a given word $x$ onto all words of $P$. Finally, the query $\formatfonc{global\text{-}max\text{-}ov()}$ is global since it evaluates all possible pairs.
Note that if $x = y$, $\formatfonc{correlation}$ returns the autocorrelation of $x$.

By definition of the Aho-Corasick automaton (AC), its structure is in fact the \vmit{trie} equipped with failure links. 
Above, we  mentioned that one can find all overlaps within the Aho-Corasick automaton of $P$ (Prop.~\ref{prop-overlap-node}) 
by traversing the chain of failure links starting at each node that 
represents a word of $P$.  However, some nodes do not represent overlaps between words of $P$. In fact, {\covi} implements a reduced Aho-Corasick automaton where precisely those nodes have been removed, and the appropriate arcs compacted (arcs corresponding to \vmit{goto} and failure links) . Hence, {\covi}'s underlying structure is not that of the AC, but a graph that we termed \emph{Extended Hierarchical Overlap Graph}. The latter is a variant of the Hierarchical Overlap Graph defined in \cite{CCR16}, except that its node set is $\{P \cup Ov^+(P)\}$ (instead of containing only the union $P$ with all maximal overlaps). In fact, this digraph has two types of arcs: those being the contractions of arcs from the \vmit{trie}, and those corresponding to failure links in the AC. Somehow, the underlying structure is a tree~\cite{Ukkonen-linear-greedy-1990}.

Experiments of Section~\ref{sec:exp} show that, because of this reduction, {\covi} can contain substantially less nodes than the full AC.

\subsection{Construction}
\label{sec-const}

To construct {\COvI}, we first build a succinct Aho-Corasick automaton, and then remove all nodes that do not correspond to an overlap between words of $P$, and finally compact the arcs between remaining nodes. Given a set of words $P$, our construction algorithm is decomposed in four steps: 1) \textbf{build the trie} $T$ of $P$, 2) \textbf{compute the failure links} of nodes of $T$, 3) \textbf{mark the nodes corresponding to overlaps} in a array $M$ and 4) \textbf{build a new structure} on the nodes in $M$.

\paragraph{Building the trie}
We assume that $P$ is given as a single text file, where the words are concatenated and separated by a special \vmit{end-of-word} symbol. 
In contrast to existing implementations, we choose a minimal structure for representing the \vmit{trie}: it has three components each stored in an array. For each node, we store the letter that connects it with its parent, a bit indicating if the node is a leaf, and a pointer to its right neighbour (if it exists). 
The neighbour array gives the position (in the array) of the 
neighbour of the current node, or $0$ otherwise. Additionally, the 
first child of each node must be in the adjacent array position of 
the current node (unless it is a leaf).

\paragraph{Computing failure links}
Using $T$, which is trie of $P$, we compute the \vmit{failure} 
links following the original algorithm of Aho and Corasick, which visits the nodes in level-wise order~\cite{AC75}.

\paragraph{Marking the nodes corresponding to overlaps}
All the overlaps between the words of $P$ (nodes that belong to the 
{\COvI}) can be obtained by traversing the chain of failure links of each node that 
represents a word of $P$ in the trie $T$ until reaching the root.
We start by creating an array $M$ to mark the nodes of $T$. After that, for each node that 
represents a word of $P$, we mark this node in $M$ and we traverse its chain of failure links until reaching either a marked node or the root. Finally, it marks all nodes of {\COvI} and only these.

\paragraph{Building the new structure}
The last step of our algorithm consists in computing the components used to represent the {\COvI} of $P$. We traverse the neighbour array of the \vmit{trie} (with nodes in depth-first order), and encode in a {\LOUDS} array the new tree with the marked nodes (with nodes in breadth-first order) (see Fig.~\ref{fig:opera:traverse}). In addition, we store in an array \formatfonc{Failure\text{-}Link} the new failure links only for marked nodes of $T$. Note that the failure link of a marked node necessarily  points to another marked node. Finally in the array \formatfonc{Depths}, each position gives the depth of a node represented by this position in $T$ (\ie\ the length of the word corresponding to this node).

%

\subsection{Supporting {\COvI} Queries}

To access easily to a node of {\COvI} representing a word of $P$, we store an array $L$ corresponding to the mapping from the words of $P$ to the nodes of {\COvI}.
Here, we give the algorithm for each query and its complexity. Let  $m$ denote the number of nodes in {\covi}.

\begin{figure}[t]
  \begin{center}
    \subfloat[]{\includegraphics[scale=0.17]{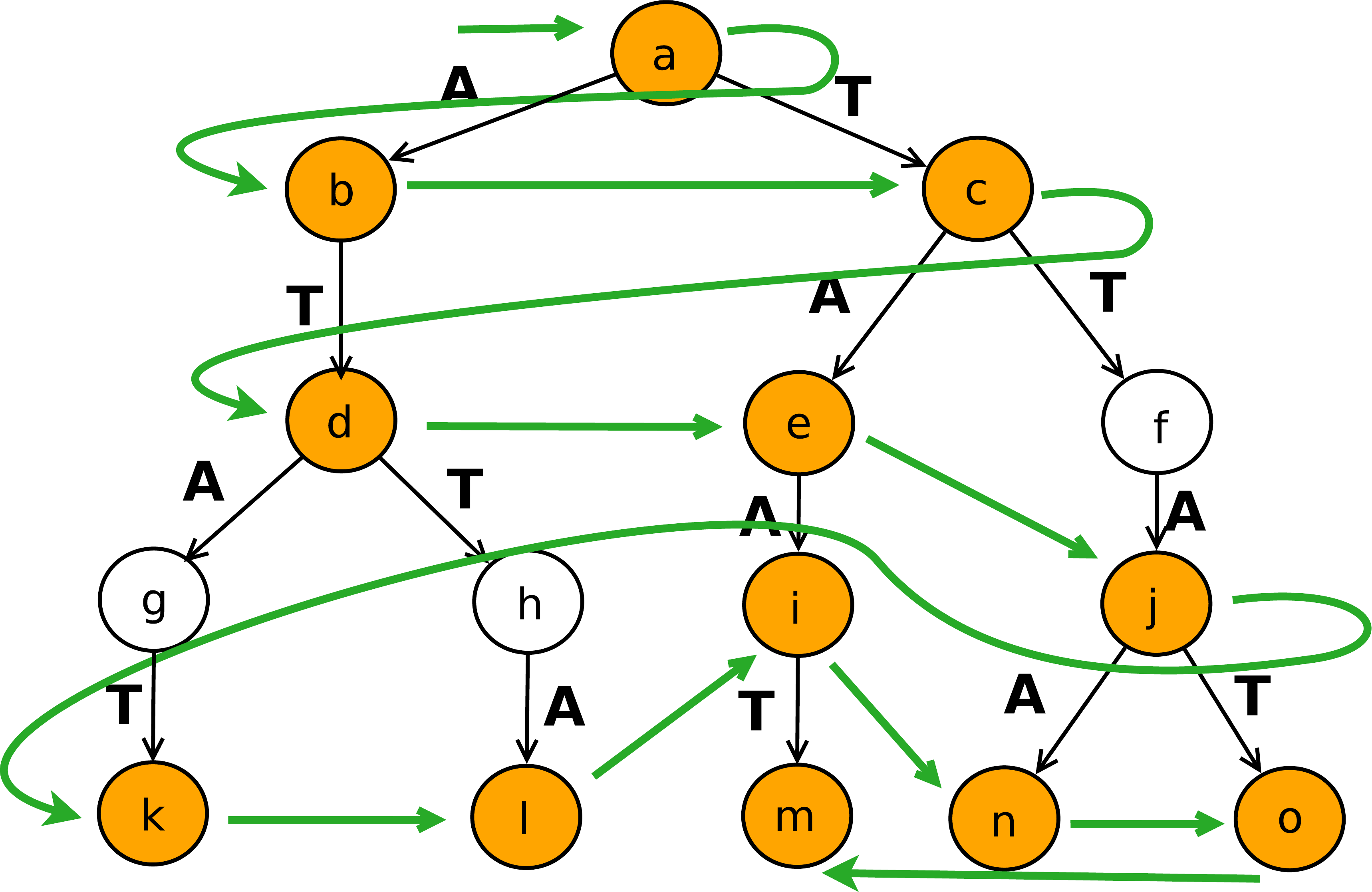}\label{fig:opera:traverse}}
    \subfloat[]{\includegraphics[scale=0.17]{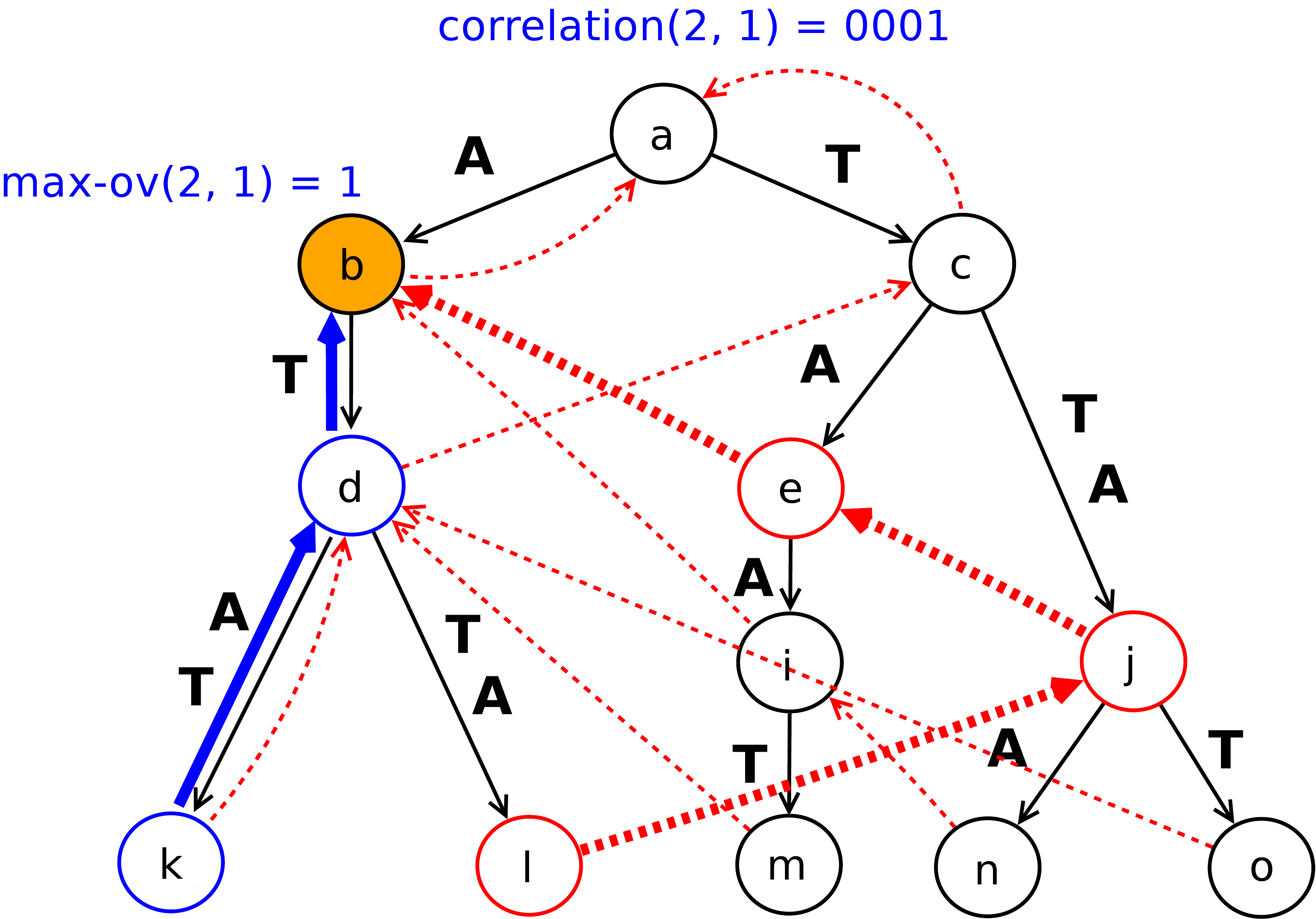}\label{fig:opera:ehog}}

  \end{center}
  \caption{\protect\subref{fig:opera:traverse} shows how we traverse the trie to create the {\LOUDS} array. Only the coloured nodes will be kept in {\covi}. \protect\subref{fig:opera:ehog}: {\covi} of $P$ and example of the $\formatfonc{max\text{-}ov}(x,y)$ and the $\formatfonc{correlation}(x,y)$ queries for $x = ATTA$ and $y = ATAT$ ($x$ is the second word of $P$, $y$ is the first). Starting from leaves \texttt{k} and \texttt{l}, the first node at the intersection of blue and dotted red paths is node $b$, whose depth equals one. The backward path of goto transitions is in bold blue, while the failure links is in dotted bold red arcs.}
  \label{fig:opera}
\end{figure}

\paragraph{Query $\formatfonc{max\text{-}ov}(x,y)$} \textbf{Complexity in $O(\lgr{x}+\lgr{y})$} \\
To compute the maximum right overlap between $x$ and $y$, we need to find the deepest node on the 
tree such that it is an ancestor of the node representing $y$ and, at the same time, it can be 
reached using iteratively the \vmit{failure} links path starting from the node representing $x$. 
In the worst case, this takes $\lgr{x}+\lgr{y}$ node queries. In practice, we access the nodes $n_x$ and 
$n_y$ representing $x$ and $y$ respectively, and update $n_x$ by $\formatfonc{Failure\text{-}Link}[n_x]$ and 
$n_y$ to $\formatfonc{Parent}[n_y]$ (using the {\LOUDS} array). Then, at each step, we check the depth of $n_x$ and 
$n_y$. If they have the same depth $d$, and are the same node or $d$ is zero, then return $d$. If not, 
in the case that $n_x$ is deeper than $n_y$, move $n_x$ to $\formatfonc{Failure\text{-}Link}[n_x]$, otherwise move 
$n_y$ to $\formatfonc{Parent}[n_y]$, and repeat the checking process (see Fig.~\ref{fig:opera}) until eventually reaching the root.

\paragraph{Query $\formatfonc{correlation}(x,y)$} \textbf{Complexity in $O(\lgr{x}+\lgr{y})$}\\
We use the same algorithm of $\formatfonc{max\text{-}ov}(x,y)$, but instead of stopping after finding the first overlap, we continue the process reporting each time we find a new overlap (until $n_x$ or $n_y$ arrive at the root) (see Fig.~\ref{fig:opera:ehog}). 

\paragraph{Query $\formatfonc{all\text{-}right\text{-}ov}(x)$} \textbf{Complexity in $O(m)$}\\
We use an array $O$ of size $m$ (the number of nodes of the {\COvI} tree) initialised with a predefined empty value. 
Then we access the node $n_x$ representing $x$ and compute $n_x:=\formatfonc{Failure\text{-}Link}[n_x]$.
From this point, we recursively set $O[n_x]$ equal to the depth of $n_x$ and update 
$n_x:=\formatfonc{Failure\text{-}Link}[n_x]$ until arriving at the root of the tree. 
Finally, for each $y$ a word of $P$, we move to the parent of the node $n_y$ (using 
the {\LOUDS} array), and check the value of $o:=O[n_y]$. If $o$ is not the empty value, then we 
know that $o$ is the maximum overlap between $x$ and $y$, otherwise we move $n_y$ to its parent 
and check the $O$ array again. 
Before reporting the $o$ overlap between two words, our methodology needs to assign the value $o$ to 
$O[n_y']$  for all the nodes $n_y'$ visited until that point.

\paragraph{Query $\formatfonc{all\text{-}left\text{-}ov} (y)$} \textbf{Complexity in $O(m)$}\\
This query is similar to $\formatfonc{all\text{-}right\text{-}ov}(x)$, because we have symmetrical definition of the failure and parent links.

\paragraph{Query $\formatfonc{global\text{-}max\text{-}ov()}$} \textbf{Complexity in $O(\lgr{P})$}\\
We know that every internal node of the {\COvI} tree represents an overlap 
between at least two (different or the same) words in $P$. 
Then, for each word $x$ in $P$ we move to its parent and mark that node as 
a candidate to be a maximum overlap in the set. 
After, we initialise the maximum depth $d=0$, and for each word $y$ in $P$ we move to its 
\vmit{failure} link node $n_y$ only checking the ones that are marked as candidate nodes. 
If the depth of $n_y$ is lower than $d$, we move to the next word. Otherwise, if it is equal 
to $d$, we add $y$ to the set of answers. If the depth of $n_y$ is greater than $d$, we 
erase the current set of answers, add $y$ to the set and update $d$ to the depth of $n_y$.

\paragraph{Query $\formatfonc{threshold\text{-}right\text{-}ov}(x,q)$} \textbf{Complexity in $O(m)$}\\
We can use a similar approach to $\formatfonc{all\text{-}right\text{-}ov}(x)$. The only difference is, 
that we add an extra condition where we check if the depth of $n_y$ is lower than $q$. If that is the 
case, then we stop the current search at that point and update the values of the $O[n_y']$ assigning 
then a zero value. Finally, at the end of the overlap search of each $y$ we only report the ones with 
depth greater than or equal to $q$. The algorithm for query $\formatfonc{threshold\text{-}left\text{-}ov}(x,q)$ is similar.

%


%% file: experi.tex

We assess the performance of {\COvI} and compare it with a baseline solution based on the Aho-Corasick automaton. 
Given the lack of an existing implementation, we created {\AC}, which stores the 
components of the Aho-Corasick automaton that are required for supporting the queries 
presented in Section~2. 
To be fair when comparing with {\COvI}, {\AC} only stores the topology of the 
generated \vmit{trie}, the  \vmit{failure} link of each node, and a mapping between 
$P$ and the nodes/leaves of the \vmit{trie}. 
In {\AC}, the arcs are labelled by a single symbol. Thus, we choose to store the \vmit{trie} topology in {\BP} format, which allows us to compute the \vmit{depth} of 
a node in constant time, thereby avoiding another array to store this information.

All experiments were performed on a Intel(R) Xeon(R) CPU $E5$-$2623$ at $3.00$ GHz 
and $\gb{125}$ of main memory.
The operating system was Ubuntu $14.04.1$, version $3.19.0$ -$59$-$generic$ Linux 
kernel.
Our {\COvI} and {\AC} data structures are implemented in {\method{C++11}}, using 
version $5.4.30$ of the {\method{g++}} compiler and the \vmit{sdsl} library \cite{GBMP14}.


\paragraph{Datasets}

The test data used was generated from the DNA, PROTEINS and ENGLISH 
text of size $100$ Megabytes, obtained from the Pizza\&Chili 
Corpus\footnote{\url{http://pizzachili.dcc.uchile.cl/texts.html}}.
From each of these files we created three datasets containing all the 
\vmit{k-mers} (with $k \in \{25,50,100\}$) extracted from each of the words 
stored in each text using a random, one to ten, skip step to compute the next 
\vmit{k-mer}. 
These datasets give us overlaps of different lengths between the words, 
depending on the value of $k$ (with a maximum overlap length of  $k-1$).
Before indexing, all words were alphabetically ordered and duplicates removed 
(using \vmit{sort} and \vmit{uniq} commands).  Although {\COvI} and {\AC} do 
not require this preprocessing, we opt for it because it is then simpler to 
assess the influence of the number of input words on memory usage. 
Table~\ref{tbl-data} lists the datasets we used and their stats. 

Note that for {\COvI} these datasets are not ideal:  they contain multiple 
overlaps for each word, but no duplicated words. This tends to lessen the difference in number of nodes between {\COvI} and {\AC}.
In practice this case is not expected. 
For example in Bioinformatics generally the set of words (reads) consist of multiple 
short words with multiple repetitions and overlaps, or a set of long words with no 
repetitions and only very small overlaps. 
Both previous described cases are ideal for {\COvI} given the construction explained 
in Section~\ref{sec-const}. 

\begin{table}[t]
  \centering
  \setlength{\tabcolsep}{2\tabcolsep}
   \input{./tbl/new_table_1.tex}

  \caption{Text files used in our experiments and their stats. The last three columns show 
  the numbers of nodes in the {\AC}, in  {\COvI}, and the percentage of the nodes of {\AC} 
  that are kept in {\COvI}.}
  \label{tbl-data}
 \setlength{\tabcolsep}{.5\tabcolsep}
\end{table}


\paragraph{Difference in number of nodes.}
By design,  {\COvI} differs from {\AC} in their numbers of nodes. The last three columns of Table~\ref{tbl-data} give the number of nodes of each structure, and the ratio of the number of nodes kept in {\COvI}. For all datasets,  {\COvI} keeps only $20$ to $32$ percents of the nodes of {\AC}. It also appears that the larger the $k$-mers, the fewer the nodes kept in {\COvI}. Indeed, the compression achieved by {\covi} increases with $k$, as the number of nodes in {\AC} increases with the length of input words. Naturally, this compression impacts the size of all three arrays of {\covi}.



\paragraph{Construction time and space.}

The final space occupied by each structure and the time spent to build {\COvI} and {\AC}  are reported for each dataset in Table~\ref{tbl-time}.  The user time is in seconds and the final storage space in Megabytes (Mb). 
The construction of {\AC} follows the same first two steps than for {\COvI} (see Section~\ref{sec-const}), except that the {\BP} representation can be computed during step two.
One observes that most of the computational time is spent by the first two steps (that is, calculating the full \vmit{trie} and its \vmit{failure} links). As these steps are the same for {\AC} and {\COvI}, it follows that the extra time needed to transform Aho-Corasick structure into {\COvI} amounts to less than $1\%$ of the total time.
Regarding memory consumption, {\COvI} uses in the worse case $\sim 1.2$ times the size of the dataset, while only requiring between $20\%$ to $40\%$ of the {\AC} size. 
Shall the datasets include repetitions, both the sizes {\AC} and {\COvI} would 
decrease similarly in function of the input size. This would impact the size of array $L$, which stores the mapping, but not their topology.

\begin{table}[t]
  \centering
  \setlength{\tabcolsep}{2\tabcolsep}
  \input{./tbl/new_tbl_times.tex}

  \caption{Time, in seconds, used to construct the {\AC} and {\COvI} data structures for each of the test files. Also, 
  for each data structure we presente the final storage space in Megabytes.}
  \label{tbl-time}
 \setlength{\tabcolsep}{.5\tabcolsep}
\end{table}


\paragraph{Query times.}

Finally we measured the queries time performances using {\COvI} and {\AC}. In the case of
{\AC} the queries are solved similarly as for {\COvI}. The main difference is that, depending 
of the query, while the number of \vmit{failure} link transitions is the same for both approaches, 
the number of \vmit{parent} transitions in {\AC} would increase in function to the number of nodes 
that are part of {\AC} and are not in {\COvI}.

We opted to present the times obtained only for the general queries (excluding queries 
$\formatfonc{threshold\text{-}right\text{-}ov}(x,q)$ and $\formatfonc{threshold\text{-}left\text{-}ov}(x,q)$ 
that are more difficult to compare because of the extra parameter $q$). 

For $\formatfonc{max\text{-}ov}(x,y)$ and $\formatfonc{correlation}(x,y)$, we randomly 
selected $100{,}000$ pairs of word indexes and reported the average time per query (in microseconds).
Similarly, for $\formatfonc{all\text{-}right(left)\text{-}ov}(x)$, we randomly selected $100{,}000$ word 
indexes and also reported the average time per query (in seconds). 
Last, we tested $\formatfonc{global\text{-}max\text{-}ov()}$, by running this query 
$1{,}000$ times and reporting the average time obtained (in seconds). Table~\ref{tbl-ov1} 
shows the results obtained for each of these queries.

\begin{table}[t]
  \centering
  \input{./tbl/new_tbl_ov_1.tex}

  \caption{Average times to perform five queries defined in Section~2. The times for the first two queries are in microseconds, the others are in seconds.}
  \label{tbl-ov1}
\end{table}

From Table~\ref{tbl-ov1} one sees that query times for $\formatfonc{max\text{-}ov}(x,y)$ and 
$\formatfonc{correlation}(x,y)$ are similar: they depend on the words' length. For these queries 
{\COvI} is always between $2$ to $4$ times faster than {\AC}. The times obtained are around $10$ fold the time for computing directly the overlap between two words, because of the use of {\LOUDS} and {\BP}, 
respectively, to move within the tree. The advantage of keeping {\COvI} in this case is that, in general finding 
and extracting the two words to be compared from a plain text takes longer than using directly our data structure.

For the queries $\formatfonc{all\text{-}right(left)\text{-}ov}(x)$ {\COvI} also performs better than {\AC} 
being $2$ to $4$ times faster than {\AC}. Notice that a naive approach would compute the right(left) 
overlaps of $x$ against each of the words in a set: it would take $p$ times the time of computing one overlap. We can 
deduce (given the number of words in each set) that our method improves over a naive approach by taking $\simeq 20$ to $30$ percents of its time. Moreover, given our approach, if the set contains repetitions, {\COvI} 
would only compute the overlaps once, while a naive approach would recompute these multiple times.

An advantage of {\COvI} is that it contains all overlap information without the unnecessary nodes of {\AC}. This avoids 
recomputing overlaps that are shared between different pairs of words. A clear proof of that, is 
the performance displayed for $\formatfonc{global}$-$\formatfonc{max\text{-}ov()}$. While a naive approach would require 
$p^2$ overlap computations, {\COvI} only uses $2p$ queries. Notice that for our datasets, computing 
this query naively would be prohibitive (taking on the order of weeks to finish), while {\COvI} takes 
only a few seconds.

\paragraph{Scalability} To probe the scalability and efficiency of {\covi} on real genomic data, we compared on a 
larger server {\covi} and {\AC} on a set of 49 million reads of 75 nucleotides each ($\simeq$ 3.5 gigabytes Gb of 
sequences; available at {\small\url{ftp://pbil.univ-lyon1.fr/pub/logiciel/kissplice/TWAS_paper/GeuvadisFastq.tar.gz}}).
{\AC} had 1.392 million nodes and occupied 5.74 Gb of memory,  while {\covi} stored 334 million nodes in 1.74 Gb, 
which is circa half the original input size. {\covi} scales up and offers even higher compression on genomic data 
than on our benchmark datasets.



%% file: tbl/new_table_1.tex
\small
\begin{tabular}{lcc cc cc cc | cc cc c}
\toprule
\multirow{2}{*}{File Name}	& \multirow{2}{*}{k-mer}     	&& \multirow{2}{*}{$\sigma$} 	&& Size		&& Number 	 &&\multicolumn{4}{c}{Thousands of nodes} 	 & Ratio (\%) \\
				&  				&&                        	&& (Megabytes)	&& of Words      && \D     {\AC}	&&     {\COvI}   &&   
\\	
\midrule
\fyle{DNA}			& \D$25$ && \D$16$  && \D{ }$447$	&&	$18{,}028{,}835$	&& \D\D{ }$248{,}356$	&& \D$70{,}357$	&& $28$
\\
\fyle{DNA}			& \D$50$ && \D$16$  && \D{ }$916$	&&	$18{,}844{,}684$	&& \D\D{ }$713{,}491$	&& $159{,}625$	&& $22$
\\
\fyle{DNA}			& $100$  && \D$16$  && $1{,}830$	&&	$19{,}000{,}429$	&& \D$1{,}661{,}261$	&& $335{,}429$	&& $20$
\\
\midrule
\fyle{PROTEINS}			& \D$25$ && \D$27$  && \D{ }$371$	&&	$14{,}975{,}787$	&& \D\D{ }$304{,}470$	&& \D$89{,}581$	&& $29$
\\
\fyle{PROTEINS}			& \D$50$ && \D$27$  && \D{ }$685$	&&	$14{,}090{,}763$	&& \D\D{ }$628{,}268$	&& $161{,}891$	&& $25$
\\		
\fyle{PROTEINS}			& $100$ && \D$27$  && $1{,}147$	&&	$11{,}908{,}565$	&& \D$1{,}111{,}355$	&& $242{,}678$	&& $21$
\\
\midrule
\fyle{ENGLISH}			& \D$25$ &&  $239$  && \D{ }$438$	&&	$17{,}692{,}377$	&& \D\D{ }$288{,}569$	&& \D$92{,}402$	&& $32$
\\
\fyle{ENGLISH}			& \D$50$ &&  $239$  && \D{ }$868$	&&	$17{,}864{,}478$	&& \D\D{ }$734{,}042$	&& $200{,}620$	&& $27$
\\
\fyle{ENGLISH}			& $100$ &&  $239$  && $1{,}729$	&&	$17{,}960{,}037$	&& \D$1{,}630{,}202$	&& $415{,}286$	&& $25$
\\	
\bottomrule
\end{tabular}

%% file: tbl/new_tbl_times.tex
\small
\begin{tabular}{l  c ccccc c ccccc c ccccc}
\toprule
File Name		&& \multicolumn{5}{c}{\fyle{DNA}} && \multicolumn{5}{c}{\fyle{PROTEINS}} && \multicolumn{5}{c}{\fyle{ENGLISH}}
\\ 
\vmit{k-mer}  		&& $25$ && $50$ && $100$ && $25$ && $50$ && $100$ && $25$ && $50$ && $100$
\\
\midrule
Construction  	
\\
Time (sec.)
\\
\hspace{0.3cm}{\AC}		&& $152$ 	&& \D{ }$304$ && \D{ }$620$ &
					& \D{ }$250$	&& \D{ }$352$	&& \D{ }$500$	&
					& \D{ }$413$	&& \D{ }$587$	&& \D{ }$998$	
\\
\hspace{0.3cm}\COvI			&& $163$ 	&& \D{ }$337$ && \D{ }$667$ &
					& \D{ }$265$	&& \D{ }$360$	&& \D{ }$551$	&
					& \D{ }$434$	&& \D{ }$638$	&& $1{,}033$	
\\
\midrule
Space 	
\\
(Megabytes)
\\
\hspace{0.3cm}{\AC}		&& $966$ 	&& $2{,}837$ 	&& $6{,}714$ &
					& $1{,}198$	&& $2{,}489$	&& $4{,}490$ &
					& $1{,}148$	&& $2{,}913$	&& $6{,}586$	
\\
\hspace{0.3cm}\COvI			&& $357$ 	&& \D{ }$797$ && $1{,}686$ &
					& \D{ }$439$	&& \D{ }$791$	&& $1{,}212$ &
					& \D{ }$460$	&& \D{ }$981$	&& $2{,}067$	
\\
\bottomrule
\end{tabular}

%% file: tbl/new_tbl_ov_1.tex
\small

\begin{tabular}{l  c ccccc c ccccc c ccccc}
\toprule
File Name		&& \multicolumn{5}{c}{\fyle{DNA}} && \multicolumn{5}{c}{\fyle{PROTEINS}} && \multicolumn{5}{c}{\fyle{ENGLISH}}
\\ 
\vmit{k-mer} 	 	&& $25$ && $50$ && $100$ && $25$ && $50$ && $100$ && $25$ && $50$ && $100$
\\
\midrule
Time (microseconds)
\vspace{0.05cm}\\ 
$\formatfonc{max\text{-}ov}(x,y)$  	
\\
\hspace{0.3cm}\AC			&& $16.90$	&& $17.71$	&& \D$37.49$ &
					& $12.33$	&& $14.06$	&& \D$22.09$ &
					& $13.50$	&& $16.88$	&& \D$26.69$	
\\
\hspace{0.3cm}\COvI			&& \D$3.38$ 	&& \D$5.26$	&& \D\D$9.25$ &
					& \D$3.11$	&& \D$6.38$	&& \D$11.11$ &
 					& \D$4.47$	&& \D$7.63$	&& \D$11.92$	
\\
$\formatfonc{correlation}(x,y)$
\\
\hspace{0.3cm}\AC		&& $16.71$	&& $17.96$	&& \D$36.87$ &
					& $12.25$	&& $14.02$	&& \D$21.73$ &
					& $13.46$	&& $16.98$	&& \D$26.53$	
\\
\hspace{0.3cm}\COvI			&& \D$3.43$ 	&& \D$5.36$	&& \D\D$9.34$ &
					& \D$3.17$	&& \D$6.42$	&& \D$11.20$ &
					& \D$4.55$	&& \D$7.72$	&& \D$11.93$	
\\
\midrule
Time (seconds)
\vspace{0.05cm}\\ 
$\formatfonc{all\text{-}right\text{-}ov}(x)$
\\
\hspace{0.3cm}\AC		&& $17.65$	&& $53.12$	&& $126.95$ &
					& $22.31$	&& $47.34$	&& \D$85.40$ &
					& $22.00$	&& $55.54$	&& $126.54$	
\\
\hspace{0.3cm}\COvI			&& \D$6.55 $ 	&& $14.52$	&& \D$31.87$ &
					& \D$8.09$	&& $15.12$	&& \D$23.78$ &
					& \D$9.10$	&& $20.17$	&& \D$43.68$	
\\
$\formatfonc{all\text{-}left\text{-}ov} (y)$
\\
\hspace{0.3cm}\AC		&& $22.21$	&& $62.42$	&& $151.04$ &
					& $30.91$	&& $64.00$	&& $101.98$ &
					& $30.12$	&& $81.44$	&& $182.89$	
\\
\hspace{0.3cm}\COvI			&& \D$8.46$ 	&& $19.27$	&& \D$41.21$ &
					& $10.55$	&& $19.46$	&& \D$29.74$ &
					& $11.05$	&& $24.31$	&& \D$51.84$	
\\
$\formatfonc{global\text{-}max\text{-}ov()}$
\\
\hspace{0.3cm}\AC		&& \D$8.07$	&& \D$9.84$	&& \D$10.70$ &
					& \D$7.39$	&& \D$7.62$	&& \D\D$6.63$ &
					& \D$8.16$	&& \D$9.64$	&& \D$10.56$	
\\
\hspace{0.3cm}\COvI			&& \D$2.16$ 	&& \D$2.59$	&& \D\D$2.93$ &
					& \D$2.08$	&& \D$2.30$	&& \D\D$2.22$ &
					& \D$2.52$	&& \D$2.92$	&& \D\D$3.30$	
\\
\bottomrule
\end{tabular}